\documentclass[aps, prx,twocolumn,longbibliography,superscriptaddress,amsmath,amssymb,floatfix]{revtex4}
\usepackage[latin9]{inputenc}
\usepackage{amssymb}
\usepackage{graphicx}
\usepackage{amsmath}
\usepackage{color}
\usepackage{mathrsfs}
\usepackage{float}
\usepackage{indentfirst}
\usepackage{mathrsfs}
\usepackage{float}
\usepackage{indentfirst}
\usepackage{textcomp}
\usepackage{comment}
\usepackage{mathtools}
\usepackage{natbib,hyperref}

\usepackage{soul}                                   

\newcommand{\COMMENTED}[1]{}

\begin{document}

\title{Stripes and spin-density waves in the doped two-dimensional Hubbard model: \\
ground state phase diagram}

\author{Hao Xu} \thanks{hxu10@email.wm.edu}
\affiliation{Department of Physics, College of William and Mary, Williamsburg, Virginia 23187, USA}
\author{Hao Shi} \thanks{boruoshihao@gmail.com}
\affiliation{Department of Physics and Astronomy, University of Delaware, Newark, Delaware 19716, USA} 
\author{Ettore Vitali} \thanks{evitali@mail.fresnostate.edu}
\affiliation{Department of Physics, California State University Fresno, Fresno, California 93740, USA}
\author{Mingpu Qin} \thanks{qinmingpu@sjtu.edu.cn}
\affiliation{Key Laboratory of Artificial Structures and Quantum Control (Ministry of Education),  School of Physics and Astronomy, Shanghai Jiao Tong University, Shanghai 200240, China}
\author{Shiwei Zhang} \thanks{szhang@flatironinstitute.org}
\affiliation{Center for Computational Quantum Physics, Flatiron Institute, New York, NY 10010, USA}


\begin{abstract}

We determine the spin and charge orders in the
 ground state 
 of the doped two-dimensional (2D) Hubbard model
 in its simplest form, namely with only nearest-neighbor hopping and on-site repulsion.
 At half-filling, the ground state 
is known to be an anti-ferromagnetic Mott
insulator. Doping Mott insulators is 
believed to be 
relevant to the superconductivity observed in cuprates.
A variety of candidates have been proposed for the ground state of the doped 2D Hubbard model. A recent work employing a combination of 
several 
state-of-the-art numerical many-body methods  
established the stripe order as the ground state 
near $1/8$ doping at strong interactions.
In this work, we apply one of these methods, the cutting-edge constrained-path auxiliary field quantum Monte
Carlo (AFQMC) method with self-consistently optimized gauge constraints, 
to systematically study the model as a function of doping and interaction strength. 
With careful finite size scaling based on large-scale computations, we map out the ground state phase diagram 
in terms of 
its spin and charge order. 
We find that modulated antiferromagnetic order 
persists  from near half-filling to about $1/5$ doping. 
At lower interaction strengths or larger doping, these ordered states are best described as spin-density waves, with essentially delocalized holes and modest oscillations in charge correlations. When the charge correlations are stronger (large interaction or small doping), they are best described as stripe states, with the holes more localized near the node in the antiferromagnetic spin order. In both cases, we find that the wavelength in the charge correlations is consistent with so-called filled stripes.    
\end{abstract}


\maketitle

\section{Introduction}
\label{intro}


The Hubbard model \cite{Hubbard238} 
is one of the most studied quantum many-body systems in condensed matter physics. 
With a very simple form, it plays a crucial role
in the exploration of correlation effects in electronic systems. It is a ``paradigmatic'' model in the realm of condensed matter physics, like the Ising model in statistical physics.
The Hubbard model hosts rich physics with the variation of interaction strength, doping level, temperature, and lattice geometry \cite{2021arXiv210400064Q,2021arXiv210312097A}. On a square lattice,
the doped Hubbard model is widely believed to be relevant to high-Tc superconductivity in cuprates \cite{RevModPhys.84.1383}. Despite the formal simplicity of the Hamiltonian,  the Hubbard model cannot be solved analytically,
except for a few special cases in the parameter 
space \cite{PhysRevLett.20.1445,PhysRev.147.392}.
Numerical methods play a key role in the study of the Hubbard model \cite{PhysRevX.5.041041,2021arXiv210400064Q}.

At half-filling, it is now established that the Hubbard model has an anti-ferromagnetic Mott insulating ground state for any finite value of the interaction strength \cite{PhysRevB.94.085140}.
What happens when holes are added to the anti-ferromagnetic state may have a crucial role to 
understanding 
the mechanism of high-Tc superconductors \cite{RevModPhys.78.17}.
Previously, a variety of ground state candidates were obtained with different methods, including stripe order/spin density waves \cite{PhysRevLett.80.1272,PhysRevB.78.165101,PhysRevLett.104.116402,PhysRevLett.113.046402} and
superconductivity \cite{PhysRevB.94.195126,PhysRevLett.110.216405}.
A study in 2017 involving four state-of-the-art numerical many-body methods 
concluded that, near $1/8$ doping, a  filled (i.e., with wavelength equal to inverse doping) stripe state was the ground state.
More 
evidence since then,  from other numerical methods studying systems with sizes sufficient to accommodate the wavelength of the stripe state,  has confirmed the existence of stripe order 
\cite{10.21468/SciPostPhys.7.2.021,PhysRevB.97.045138,2021arXiv210601944M,sorella2021phase} 
in the doped Hubbard model.

The study of stripe order in the Hubbard model can be traced back to 1980s in Hartree-Fock \cite{PhysRevB.40.7391,PhysRevB.39.9749,MACHIDA1989192} and 1990s
in density matrix renormalization group (DMRG) \cite{PhysRevLett.80.1272}
calculations. 
Work to study stripes in cuprate superconductors was also performed on the two-dimensional Hubbard model using dynamical mean-field theory \cite{fleck1,fleck2}
and slave boson methods \cite{PhysRevB.73.174525}.
Constrained-path auxiliary-field quantum Monte Carlo (AFQMC) calculations
showed spin-density wave states at intermediate interaction strengths which turned into stripe states with increasing interaction, and determined the wavelength of the collective modes \cite{PhysRevLett.104.116402}.
In both the SDW and stripe states, a unidirectional order is established with the   anti-ferromagnetic correlations
displaying a $\pi$ phase flip across nodes.  In the SDW state, the hole density variation is small. In the stripe phase, the doped holes concentrate in the nodal region of the spin modulation.


Stripe order is commonly observed in cuprates \cite{doi:10.1080/00018732.2021.1935698}. There is evidence suggesting that the stripe order might be the origin of pseudo-gap phase of cuprates \cite{PhysRevX.11.031007} at finite temperatures above the superconducting transition temperature. 
As a result there exists a very large body of work which 
 focuses on modeling and understanding stripe states on different models using various analytical and numerical approaches. Even within the Hubbard model, the
 connection with cuprates is still not fully established (e.g., filled stripes vs.~half-filled stripes in real materials), and there 
 remain many questions to be addressed, for example,  many finite-temperature properties, the role of $t'$ and other terms, the presence/absence of superconductivity and its relation to stripes, etc (see, e.g., Refs.~\cite{2021arXiv210312097A,2021arXiv210400064Q} for recent reviews). 
 In this work 
 we focus on the  the ground state of the pure $(t'=0)$ Hubbard model as a fundamental model, in particular, on the nature of the magnetic and charge orders as a function of  doping and interaction strength in this model.

  The Hubbard model has presented a long-standing challenge to condensed matter physics and beyond. A major reason for this challenge is that 
  there is often such a small energy scale
separating different types of orders (filled stripes, half-filled stripes, spin-density waves, phase separation, superconductivity, etc) that the outcome depends delicately on the particular system (cluster size, supercell size and shape, boundary condition, etc) and the 
particular approximation of the applied methodology and its accuracy/convergence in the calculations. This makes reliable predictions of the ground state in the thermodynamic limit (TDL) very difficult and is reflected by the wide varieties of results (often conflicting) in the literature. As mentioned, recent work involving careful benchmark and multi-method comparisons has led to significant progress.
  In this work, we present a systematic study 
  to determine the ground state phase diagram of the doped
Hubbard model by scanning the doping level 
and interaction strength $U$, with a particular focus on spin and charge orders. 
We use the constrained path auxiliary field quantum Monte Carlo (CP AFQMC) method \cite{Zhang97,PhysRevB.78.165101} with self-consistently optimized constraint 
\cite{PhysRevB.94.235119},
 which is one of the methods used to establish 
 the stripe phase near $1/8$ doping \cite{Zheng1155}. 
Spin and hole density patterns are computed in very large supercell sizes
 to explore possible phase transitions in the system and to determine
the phase boundaries. With careful finite size scaling, we map out the
phase diagram as a function of doping level and interaction strength $U$ in the TDL. 
We find that, when the interaction is sufficiently strong, the spin-density wave and stripe orders persist from small doping near half-filling to about $1/5$ doping.
 
 The rest of the paper is organized as follows. In Sec.~\ref{method} we introduce the Hubbard model and the method we use. In Sec.~\ref{stripe} we discuss the details
 on how to determine the presence or absence of order and the wavelength of the collective mode 
 in the TDL. The phase diagram is shown in Sec.~\ref{phase}.
 We then summarize the work with a conclusion and perspective in Sec.~\ref{con}.

\section{Model and Methodology}
\label{method}

\subsection{model}
The Hamiltonian of the Hubbard model is as follows:
\begin{equation}
H = -t\sum_{\langle i,j \rangle ;\sigma} c^{\dagger}_{i\sigma} c_{j\sigma} + U\sum_{i}n_{i\uparrow}n_{i\downarrow} + \sum_{i} v_{i\sigma} n_{i\sigma} \, 
\label{1}
\end{equation}
where the coordinates of the lattice site labeled $i$ are given by ${\mathbf r_i} = (i_x,i_y)$. $c^{\dagger}_{i\sigma}$($c_{i\sigma}$) represents the creation (annihilation) operator on site $i$, 
with $\sigma=\uparrow, \downarrow$ being the spin of the election. The operator $n_{i\sigma} = c^{\dagger}_{i\sigma} c_{i\sigma}$ measures the number of electrons with spin $\sigma$ on site $i$. We set $t$ as the energy unit in this work. The first (second) term in the Hubbard model represents the kinetic (interaction) energy. The last term is a spin-dependent potential from an external field, which is applied to explicitly break the SU(2) symmetry.
We use the external field as a pinning field, applied to the edges, in such a way that the symmetry breaking allows us to measure local densities as opposed to the more demanding correlation functions \cite{PhysRevLett.99.127004}. 
As we illustrate below (Fig.~\ref{diff_h_pin}),  the effects of the pinning field are negligible in the
bulk of the system.  So the details of the field does not affect the characterization of the ground state.  
We have also tested the cases with charge pinning field alone and simultaneous spin and charge pinning field, and the results are consistent with those obtained with spin pinning field only.


To characterize the spin order, we measure the staggered spin density $S_i = \frac{1}{2} (-1)^{(i_x+i_y)}\langle n_{i\uparrow} - n_{i\downarrow} \rangle $
in the presence of the symmetry-breaking pinning field mentioned above.
To characterize the
 charge order, 
 we measure the local hole density $h_i = \langle 1 - n_{i\uparrow} - n_{i\downarrow} \rangle$. 
We denote the average hole density, or doping, in the system as $\delta = 1 - N_e / N_{site}$ where
$N_e$ is the total number of electrons and $N_{site}=L_x\times L_y$ the number of sites of the lattice.
In most calculations, a cylindrical geometry is adopted to accommodate the pinning fields. 
We study 
rectangular lattices with either open or periodic boundary conditions on the longer direction ($x$), and either periodic or twist boundary condition on the shorter one ($y$). Below when we present the results, unless otherwise specified,  the default will be cylindrical cells, namely open along x and periodic along y.
This allows us to study systems with a size which can accommodate one or multiple periods of density waves or stripe order.
We vary the aspect ratio of the simulation cells to confirm the robustness of the results, as discussed in Sec.~\ref{stripe}.


\subsection{AFQMC method and self-consistent constraint}
We employ the constrained-path auxiliary field quantum Monte Carlo (CP AFQMC) method \cite{Zhang97,PhysRevB.78.165101} to calculate the ground state of the doped 2D Hubbard
model in this work. In CP AFQMC, a trial wave-function is used to control the fermion sign problem \cite{PhysRevB.41.9301,AFQMC-lecture-notes-2019}.
In this work we employ trial wave functions of an unrestricted Hartree-Fock form. Recently, we developed a strategy to optimize the trial wave-function self-consistently to reduce the constraint error and to  minimize the 
dependence
of results on the  trial wave-function \cite{PhysRevB.94.235119}. 
 CP AFQMC has proved to be highly accurate in several key benchmarks and, with its latest algorithmic advances,  
has played an important role in the recent advances in the Hubbard model
\cite{PhysRevX.5.041041,Zheng1155,PhysRevX.10.031016}.

\begin{figure}[t]
	\includegraphics[width=80mm]{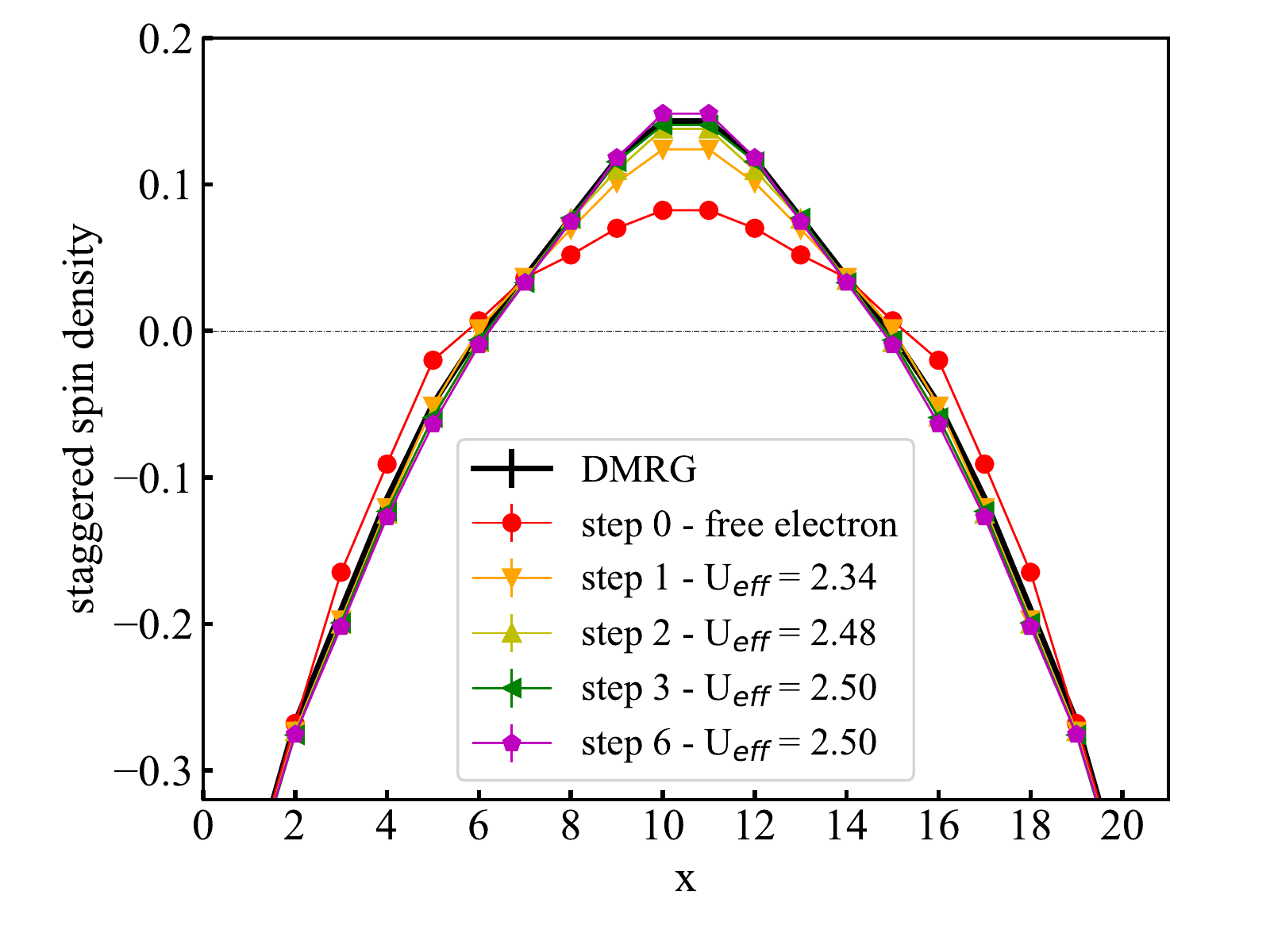}
	\includegraphics[width=100mm]{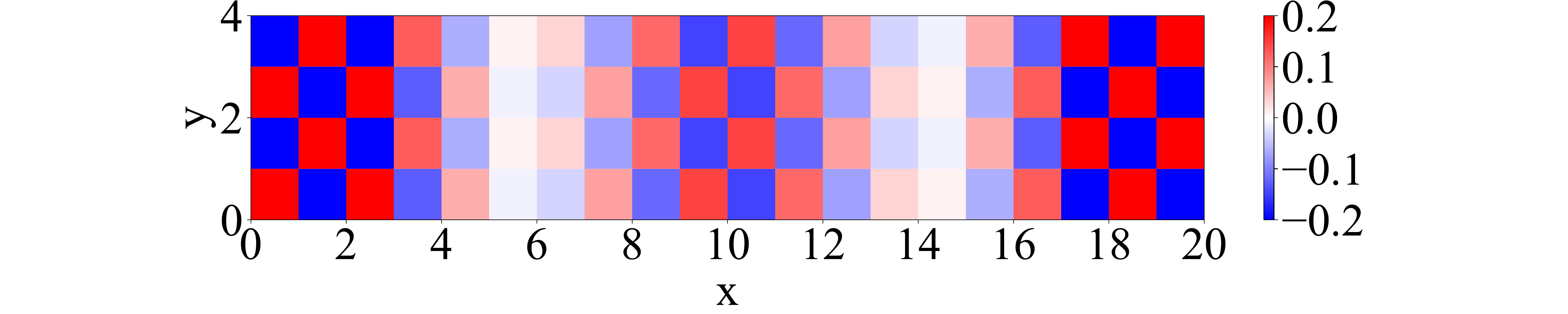}

	\caption{The spin density for $20\times4$ system with $1/10$ doping and $U = 6$. 
	Upper: staggered spin density $(-1)^{i_x+i_y}S_z (i)$ along $x$-direction  in the self-consistent CP AFQMC calculation.
	Results are averaged over different rows ($i_y$ values). 
	The converged CP AFQMC result agrees well with the accurate DMRG results (black line).  
		  Lower: a color map of the converged CP AFQMC spin density, where we can clearly see two $\pi$ phase flips. $U_{\rm eff}$ is the effective interaction
		  strength used to generate the next step trial wave-function in the self-consistent procedure \cite{PhysRevB.94.235119}.
		  }
	\label{benchmark_20_4_spin}
\end{figure}

%

\section{Determining spin and charge orders}
\label{stripe}

\subsection{Benchmark and the effect of pinning fields}

We first take the $20\times4$ cylinder with $U = 6$ and $\delta = 0.1$ as an example to 
illustrate the method and provide a sense of its procedure and accuracy.
For a narrow cylindrical 
system such as this one, 
DMRG \cite{PhysRevLett.69.2863,PhysRevB.48.10345} can provide highly accurate results for benchmark.  
A pinning field is applied at the edges of the cylinder to induce local antiferromagnetic order:
$v_{i\downarrow}=-v_{i\uparrow}=(-1)^{i_x+i_y}v_p$ for $i_x=1$ and $i_x=L_x$. The strength of the pinning field is  $v_p = 0.5$ here. 
 
\begin{figure}[t]
	\includegraphics[width=84mm]{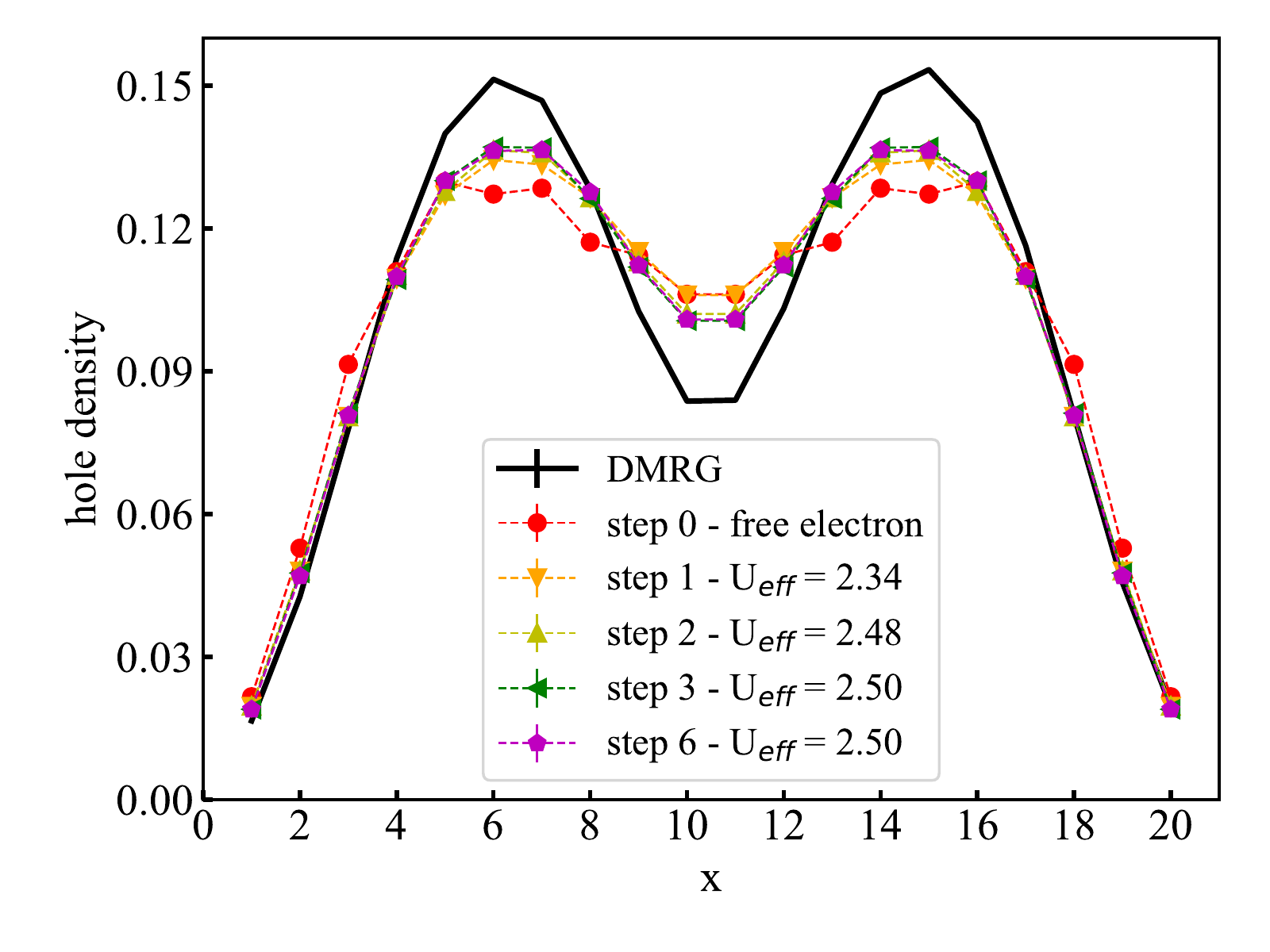}
	\caption{
	The corresponding hole density in the same system as in  Fig.~\ref{benchmark_20_4_spin}: 
	$20\times4$ cylinder with $1/10$ doping and $U = 6$. There is noticeable discrepancy in the  self-consistent CP AFQMC hole density 
		from the accurate DMRG result (black line). However the stripe structure is the same. }
	\label{benchmark_20_4_hole}
\end{figure}

The result for the staggered spin density is shown in Fig.~\ref{benchmark_20_4_spin}. 
We start the self-consistent iteration with the free-electron ($U_{\rm eff} = 0$) trial wave-function.
The energy from the free-electron trial wave-function is $-66.74(1)$, which is very close to the exact  (DMRG) energy of $-66.82(1)$. 
(Note that the energy computed from CP AFQMC with the so-called mixed estimate is not variational \citep{PhysRevB.59.12788}). However, the staggered spin density from
the free-electron trial wave-function displays some significant discrepancies with respect to the exact result as seen in Fig.~\ref{benchmark_20_4_spin}. As detailed in \cite{PhysRevB.94.235119}, we set up a self-consistent loop using the CP AFQMC solution to determine a mean-field 
solution with an effective $U$, $U_{\rm eff}$, which minimizes the difference between its density (or density matrix) and that from AFQMC. This solution (which has broken spin symmetry) is used as the new trial wave function, and the process is iterated to convergence.
After $6$ iterations, the CP AFQMC results are indistinguishable from the exact results, consistent with previous studies \cite{PhysRevB.94.235119}. The result for hole
density is shown in Fig.~\ref{benchmark_20_4_hole}. There is still noticeable 
discrepancy 
in the converged local hole densities, but the pattern is the same as the exact DMRG results.


%

\begin{figure}[b]
	\includegraphics[width=84mm] {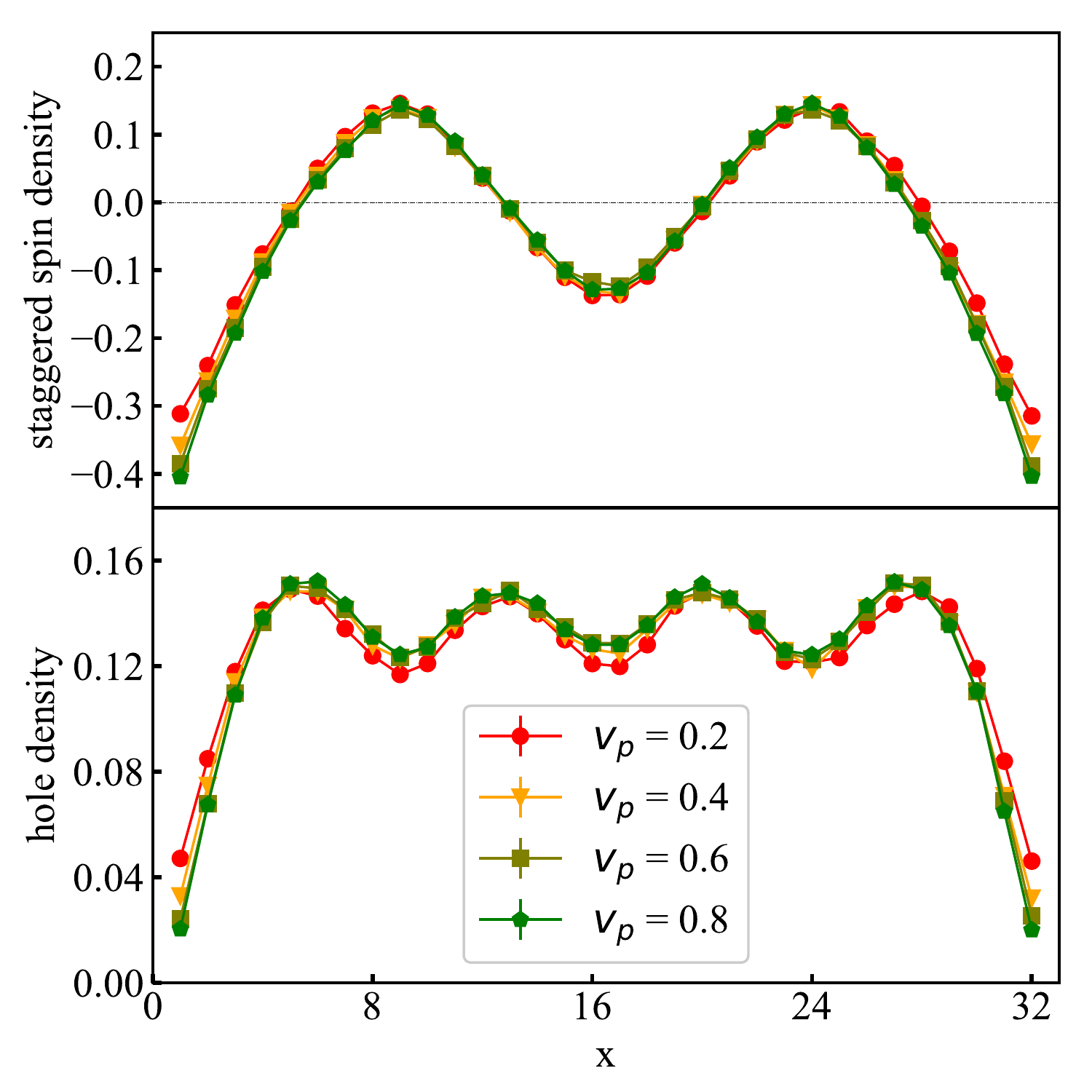}
	\caption{Insensitivity of the long-range order to the strength of the local pinning field. 
	Staggered spin density (up) and hole density (down) are shown for a $32\times6$ cylinder with $1/8$ doping and $U = 6$. 
		The strengths of the pinning field ranges from $0.2$ to $0.8$.  
		Converged results from  self-consistent CP AFQMC are shown for each system.
		The pinning field has little effect on the spin and hole density, especially in the ``bulk" of the system.}
	\label{diff_h_pin}
\end{figure}

In Fig.~\ref{diff_h_pin}, we study the effect of the strength of the pinning field on the results. The system is a $32 \times 6$ cylinder with $U = 6$ and $\delta = 1/8$.
Pinning fields are applied on the two edges, similar to the setup in the previous example. 
The strength of the pinning field, $v_p$, is now varied by a factor of $4$, from $0.2$ to $0.8$. We see 
that both the staggered spin density and hole density
remain essentially unchanged in the ``bulk" of the system. This validation shows the viability of probing long-range order with 
local pinning fields (provided that sufficiently large system sizes can be studied).

\subsection{Finite size scaling}
\label{ssec:fs-scaling}

\begin{figure}[t]
	\includegraphics[width=84mm] {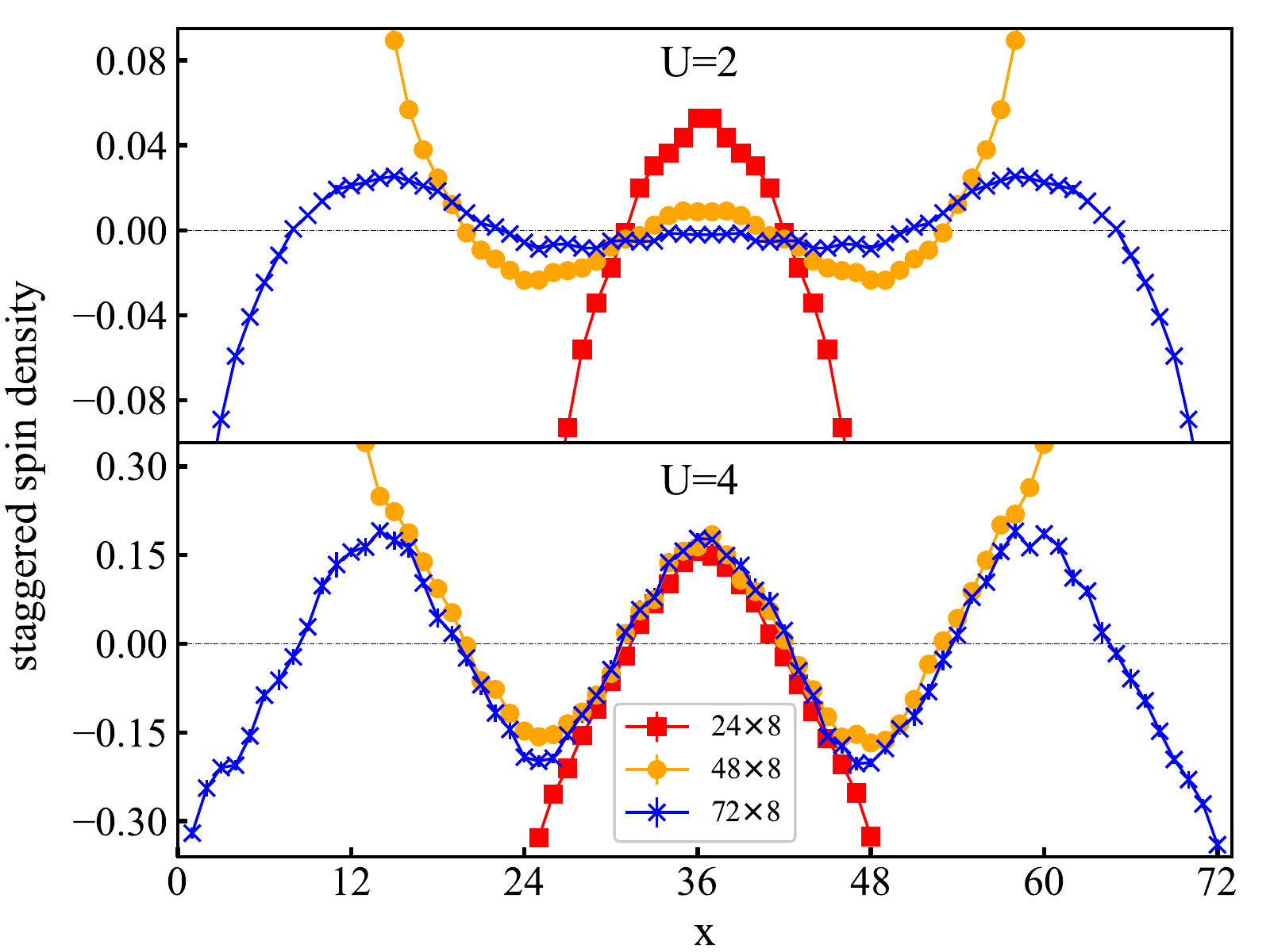}
	\caption{Presence and absence of long-range order with supercell system size. The staggered spin density is shown for two interaction strengths, 
	 $U = 2$ (top) and $U = 4$ (bottom), at $\delta=1/12$ doping, each for 
	a sequence of supercell sizes.
All results are for cylinders with
		width $8$, and the staggered spin densities are plotted along the long ($x$)  direction. 
		As the length of the cylinder is increased, the staggered spin density vanishes for $U = 2$ but remains a constant
		for $U = 4$. 
		}
	\label{U_2_4_spin}
\end{figure}

At each set of system parameters ($U$, $\delta$), we probe the order in a range of (large) lattice sizes. A true ground-state long-range order will persist with 
increasing system size, while a short-range correlation induced by the local pinning field will die out as the system size grows. 
This is shown in Fig.~\ref{U_2_4_spin}, in which the spin orders are computed in width-8 cylinders at $\delta=1/12$ doping, with $U = 2$ and $U = 4$, respectively. 
At $U=2$ the spin density in the ``bulk" of the system tends to zero as the length of the system is increased, while it is almost unchanged at $U = 4$ for $L_x$ 
from $24$ to $72$, displaying a spin-density wave (SDW) with a consistent wavelength.

\begin{figure}[t]
	\includegraphics[width=84mm] {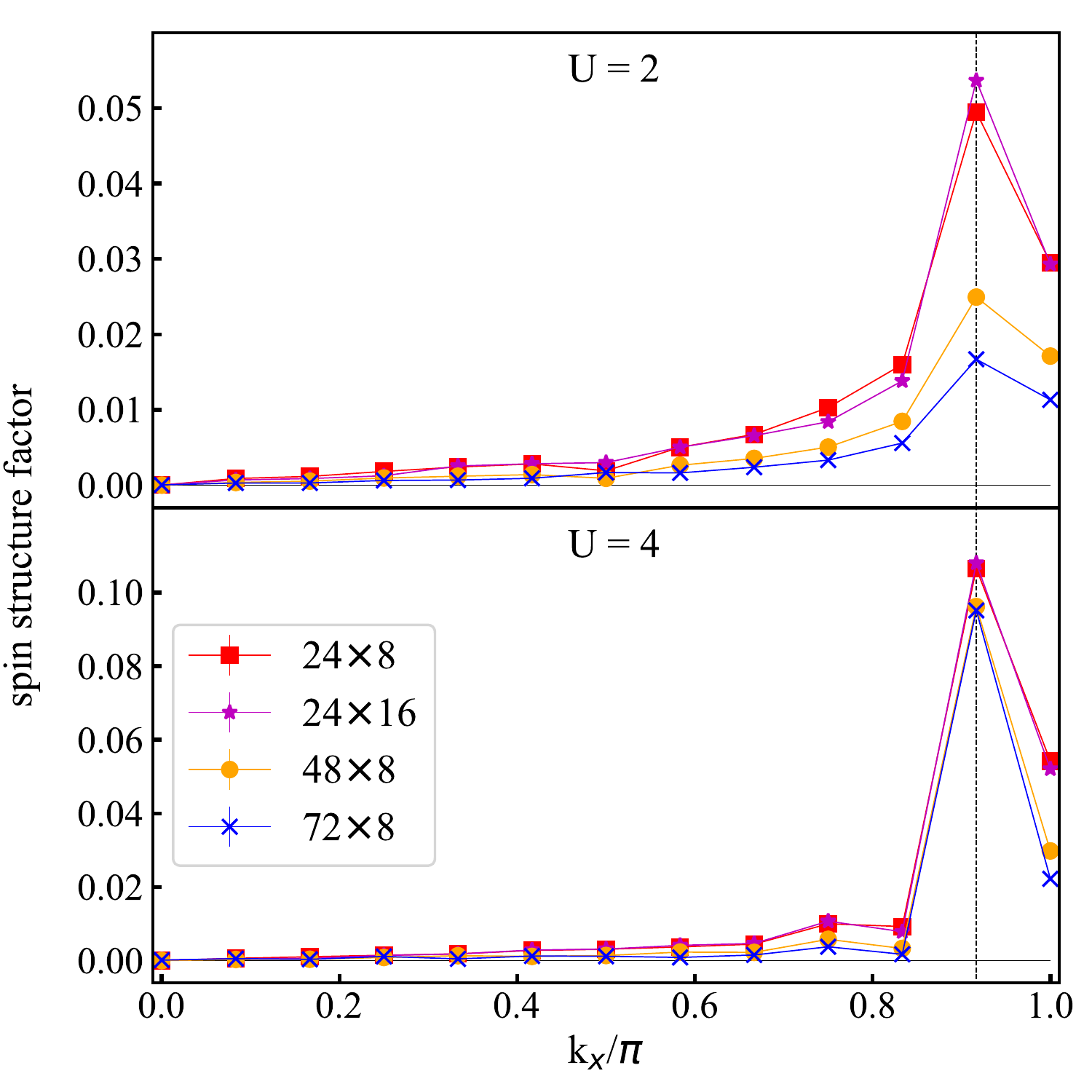}
	\caption{Spin structure factor $S_s(k_x,\pi)$ for a variety of simulation cell sizes, at two interaction strengths,   $U = 2$ (top) and $U = 4$ (bottom).
	All systems are at $\delta=1/12$ doping. A peak is seen at $k_x=\frac{11}{12} \pi$. With the  increase of $L_x$, the peak decreases and vanishes 
	at  $U = 2$ but saturates at $U=4$. Note the different vertical scales in the two panels.
	}
	\label{U_2_4_spin_struct}
\end{figure}

As a more quantitative probe of the order, we calculate the spin structure factor $S_s({\mathbf k}) = \frac{1}{N} \sum_{{\mathbf r_i}}e^{i\,{\mathbf k}\cdot{\mathbf r_i}} \langle n_{i\uparrow}-n_{i\downarrow} \rangle $, where ${\mathbf k}=(k_x, k_y)$, with $k_x=n_x\,2\pi/L_x$ and 
$k_y=n_y\,2\pi/L_y$ ($n_x\in[0,L_x)$ and
$n_y \in [0,L_y)$ are integers).
The results are shown in Fig.~\ref{U_2_4_spin_struct}.
At  $U = 4$,
a peak is seen in the spin structure factor at ${\mathbf k}_p= ((1-\delta)\pi, \pi)$, i.e., $(\frac{11}{12}\pi, \pi)$ in this case, which agrees with the wave-length of the SDW in Fig.~\ref{U_2_4_spin}.
The height of the peak saturates among the larger supercells. 
At $U = 2$, a smaller peak is also present at ${\mathbf k}_p$. 
However, the value of the peak decays as system size $L_x$ is increased.

\begin{figure*}  
        \includegraphics[width=55mm]{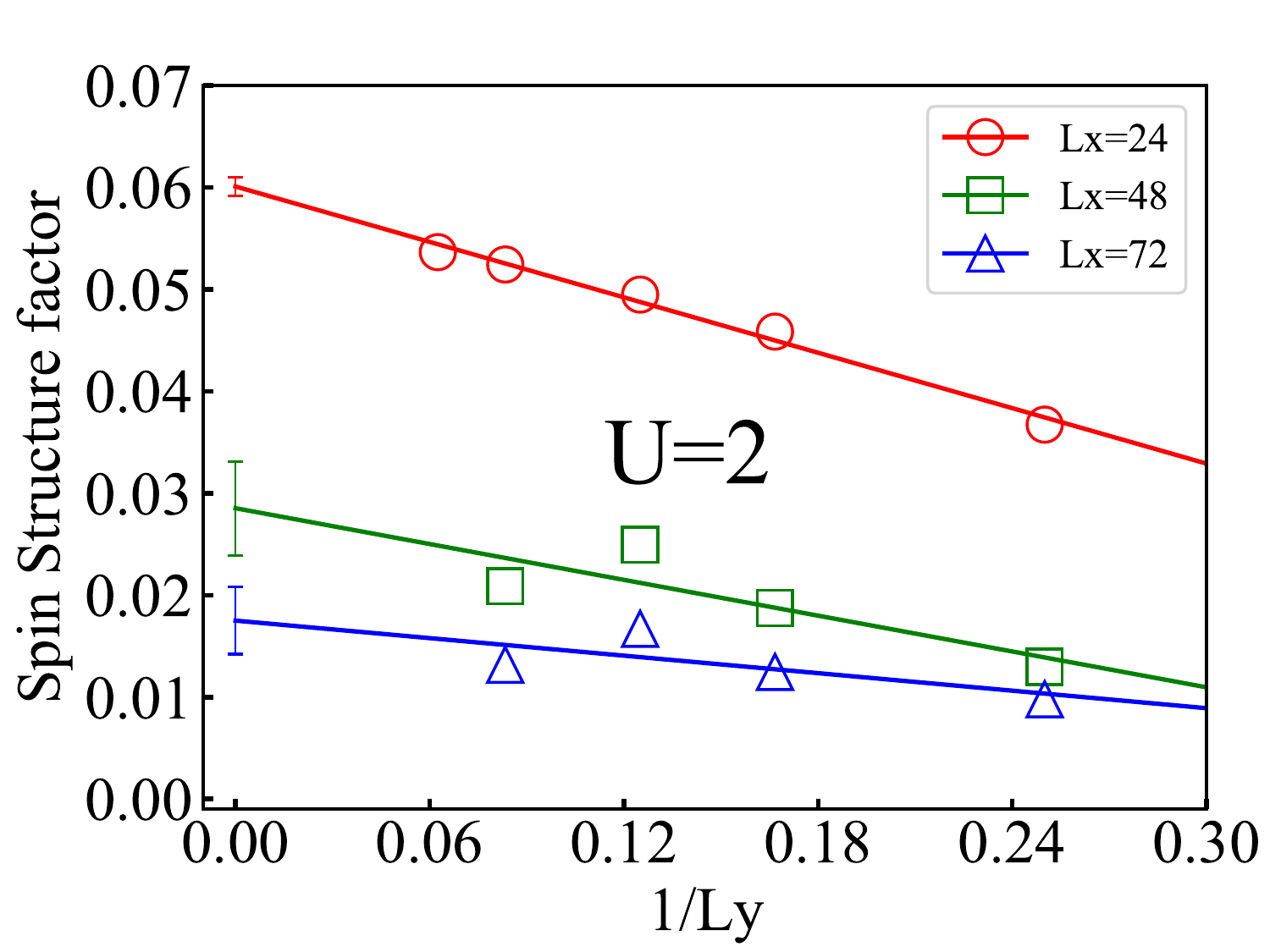}
        \includegraphics[width=55mm]{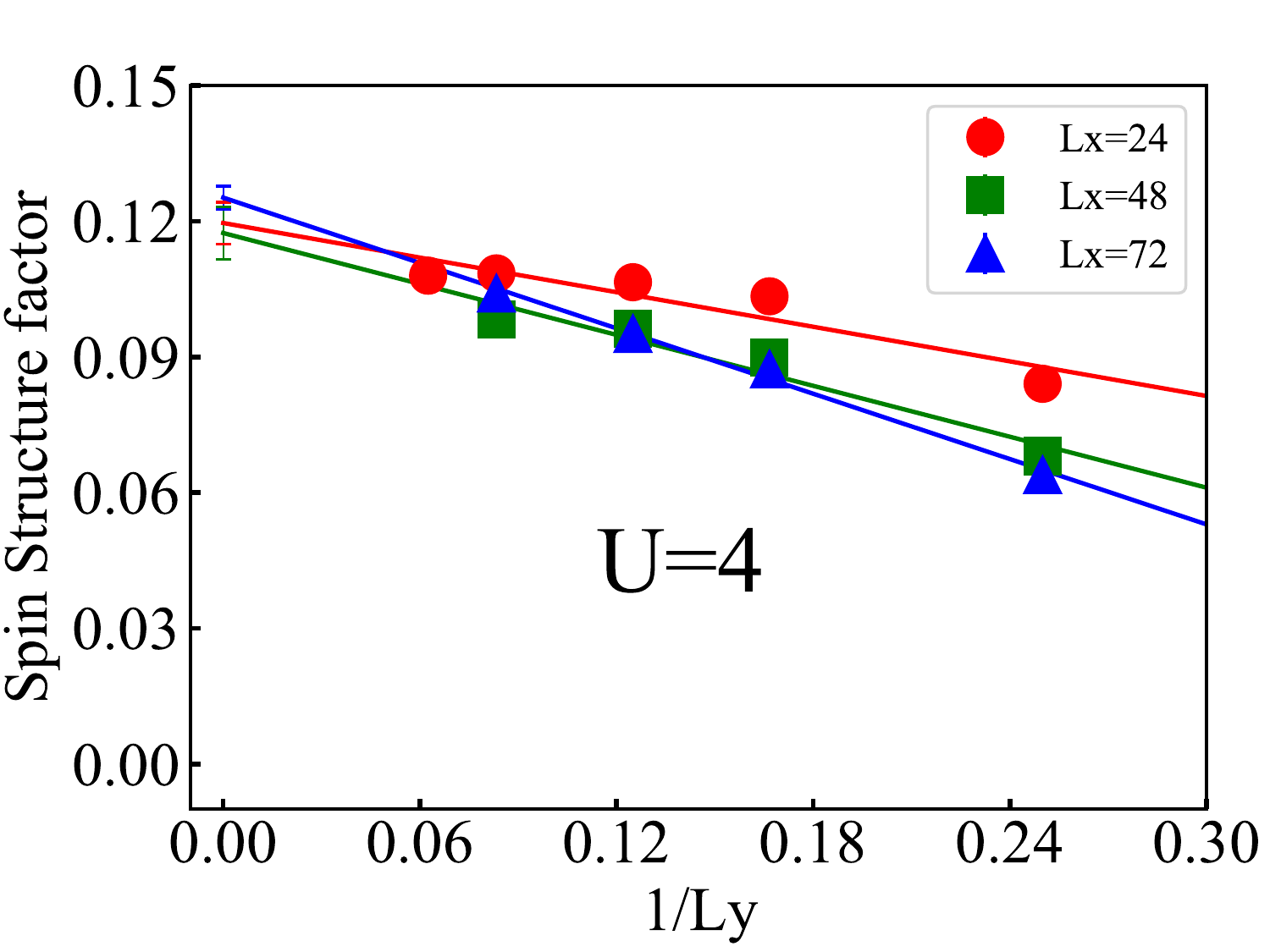}
        \includegraphics[width=55mm]{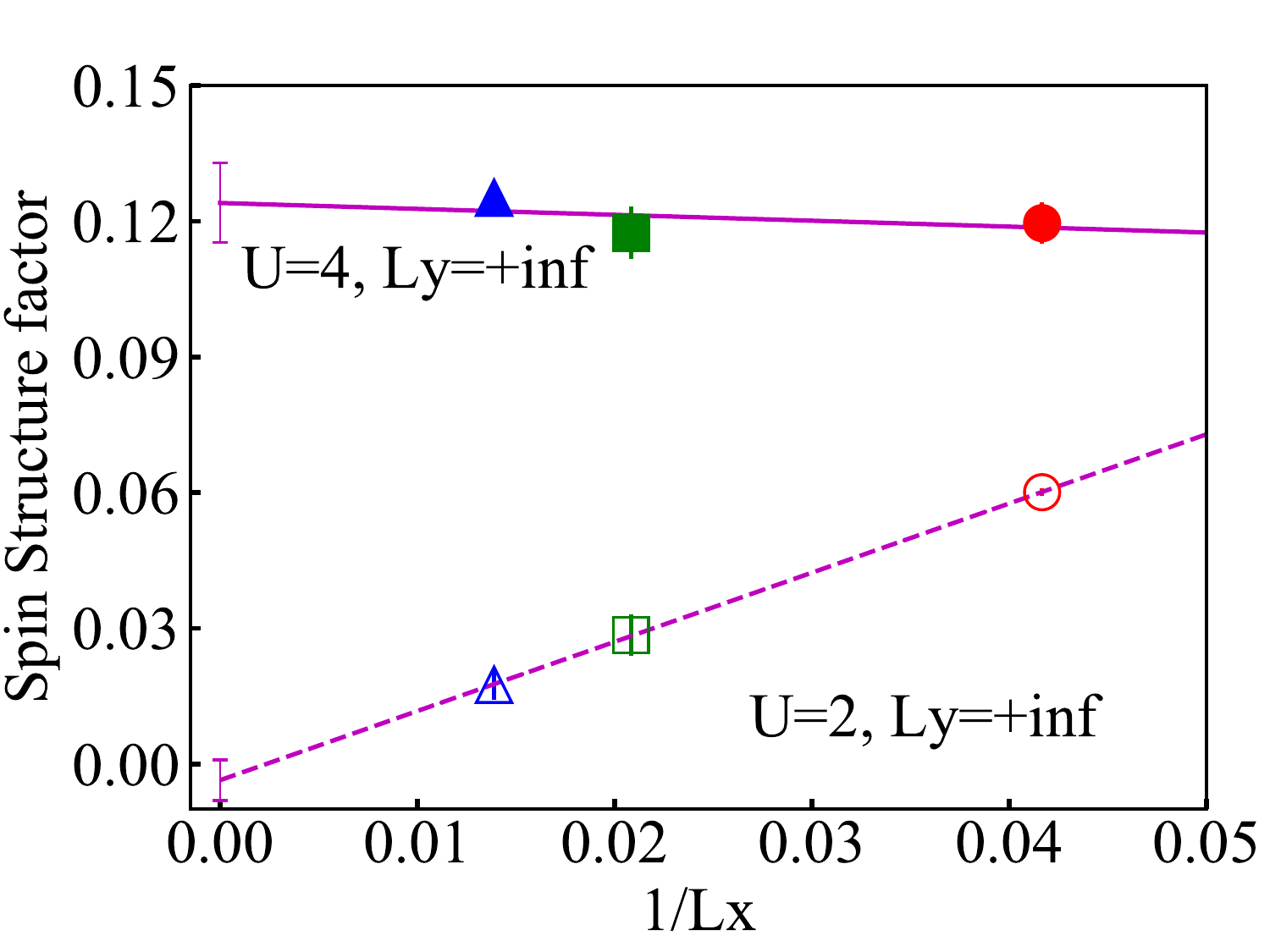}

\caption{Finite size scaling of the spin structure factors in Fig.~\ref{U_2_4_spin_struct}. 
An extrapolation with $L_y$ is performed  for each set of cylinders with the same length, shown in the left panel for $U=2$ and middle panel for $U=4$. 
This is  
followed by an extrapolation with respect to the length of the cylinders, $L_x$, shown in the right panel for both $U$ values. 
At the TDL, the peak value of the spin structure factor vanishes for $U = 2$, while it reaches 
a finite value for $U =4$. 
	}
        \label{U_2_4_tdl}
\end{figure*}

We next perform a finite size scaling of the values of the spin structure factor at ${\mathbf k}_p$ 
(the peak position). 
In order to reach the TDL, 
we extrapolate $S_s({\mathbf k}_p)$, 
first as a function of the width$(L_y)$ of the system, followed by an extrapolation as a function of the length$(L_x)$. 
This procedure is shown in Fig.~\ref{U_2_4_tdl}.
At $U = 2$ (left panel), the extrapolated values for $L_x = 24, 48$, and $72$ are $0.060(1),0.029(5)$, and $0.018(3)$ respectively,
while  
at $U = 4$ (middle panel), the corresponding values are
$0.120(5), 0.117(6)$, and $0.125(3)$. 
Extrapolations of these results with $L_x$ yield the following  values of the spin structure factor at the TDL:    
	$-0.003(5)$  at $U=2$ and $0.123(9)$ at $U=4$.
From these results we conclude that, for doping $\delta=1/12$,
 a spin 
order is absent 
at $U = 2 $ but present at $U = 4 $.  (These two points are indicated as points A and B in the
phase diagram in Fig. \ref{phase-diagram}.)
We systematically apply this procedure to determine the presence of order for each set of Hamiltonian parameter hence an estimate of the
critical interaction strength $U_c$ for the appearance of an SDW or stripe order, and map out  a phase diagram for $U\lesssim 12$. 

%


\subsection{Determining the wavelength of the collective modes}
\label{ssec:determine-lambda}

We find that a modulated spin order appears for doping values up to about $\delta \sim 1/5$, often accompanied by charge orders. 
We will further discuss the properties of these collective modes and provide a detailed phase diagram below.

\begin{figure}
     \includegraphics[width=84mm]{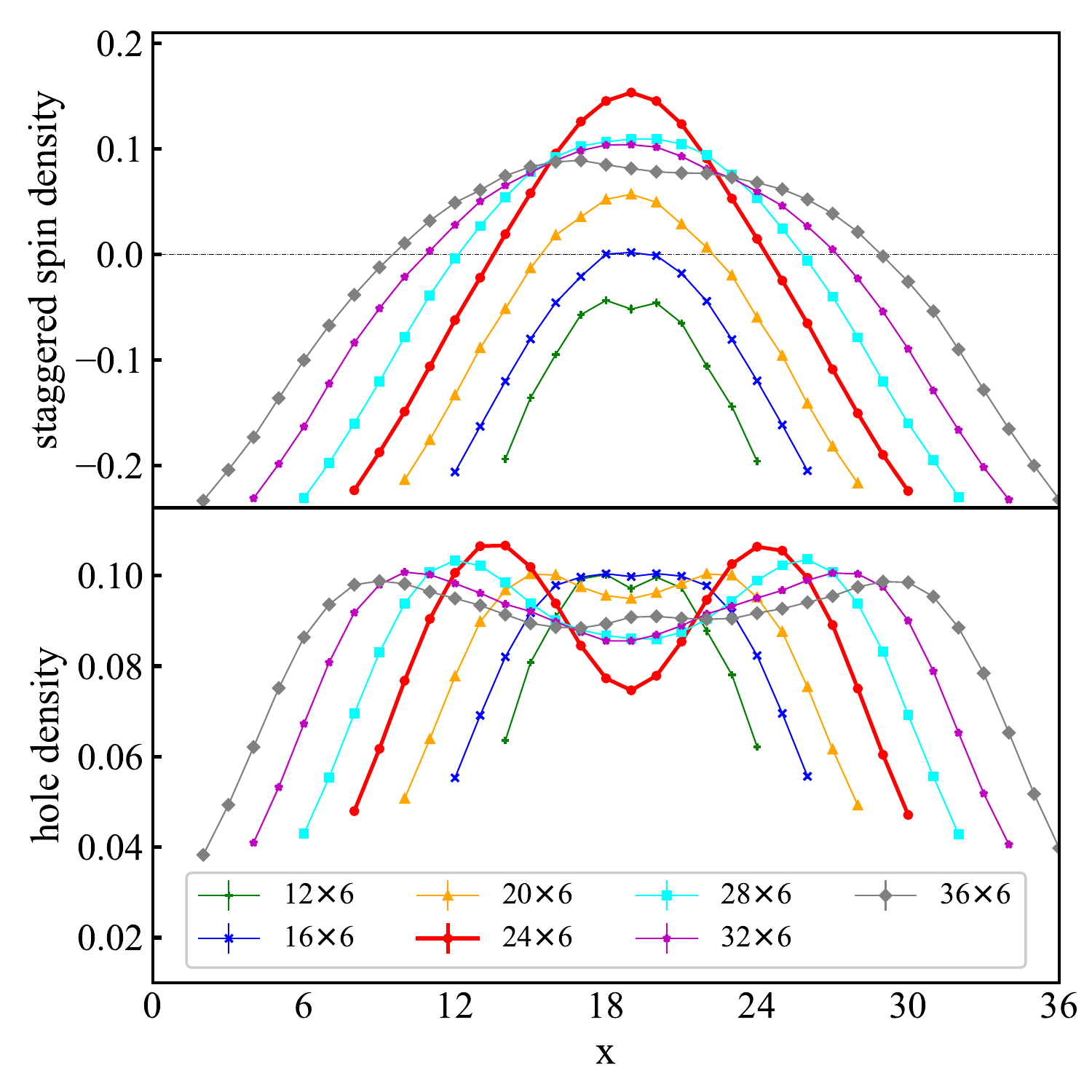}
    \caption{Staggered spin (top panel) and hole (bottom) densities at $\delta=1/12$ doping and $U=4$, in width-6 cylinders as the length $L_x$ is varied.  Results are omitted at the left edge ($i_x=1$), where the pinning field is applied. When $L_x$ is commensurate with the expected wavelengths for spin and charge orders
    ($2/\delta$ and $1/\delta$), the spin and charge 
    density waves are least frustrated and have the largest amplitude. 
}
    \label{1_12}
\end{figure}

Here we describe our investigation of 
the wavelength of the collective mode in the spin and charge order in the ground state.
We first illustrate the procedure using the case of $\delta=1/12$ and $U=4$ 
as an example.
In Fig.~\ref{1_12}, we vary the length of the cylinder (hence also $N_e$, in order to maintain the same $\delta$) 
while keeping the width fixed at $L_y=6$.  
A pinning field of strength $v_p = 0.5$ is applied only at the left edge ($i_x=1$) and periodic boundary condition (PBC) is used along x direction.   For y direction,  twist average boundary condition (TABC) is used to further reduce the finite size effects \cite{PhysRevE.64.016702}. 
From Fig.~\ref{1_12}, we see that the staggered
spin density becomes strongest, and frustration is minimized when $L_x=24$. The corresponding charge order also forms a regular wave with 
hole density peaks at the nodal position of the spin order. This is consistent with a wavelength of $2/\delta$ for the SDW and $1/\delta$ for the CDW. 
At larger interaction strengths, the SDW evolves into a stripe order \cite{PhysRevLett.104.116402},  and our results suggest that the stripes are filled.
We use this procedure, combined with finite-size scaling as discussed in Sec.~\ref{ssec:fs-scaling}, to establish the order in the TDL and determine 
its wavelength. 
More examples are shown in the appendix.

\begin{figure}
    \includegraphics[width=84mm]{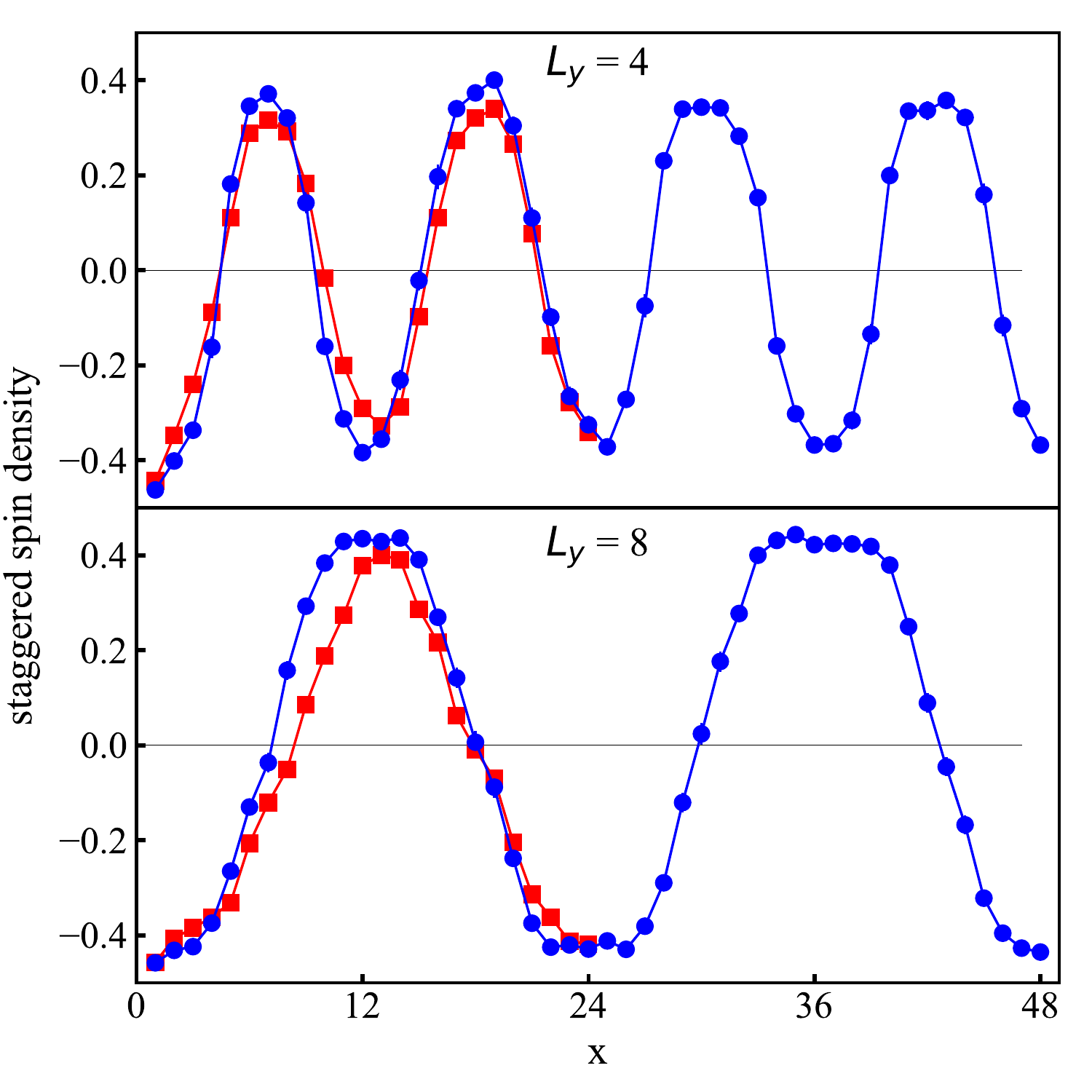}
    
    \caption{Different behaviors in width-4 (top panel) and width-8 (bottom) cylinders.
    Staggered spin densities are shown for $\delta=1/12$ doping with $U=12$. 
    The red curves show results for cylinders of length $L_x=24$, while the blue curves show those with $L_x=48$.
    }
    \label{1_12_4_8}
\end{figure}

 Recent studies from DMRG \cite{PhysRevResearch.2.033073}
 and the minimally entangled typical thermal states (METTS)  \cite{PhysRevX.11.031007} 
 methods 
 have found half-filled stripes in width-4 cylinders. For example, at $U=12$ half-filled stripes are identified as the ground state for all doping values  
 below $\sim 1/9$  \cite{PhysRevResearch.2.033073}, while from METTS $\delta=1/16$ at $U=10$ is seen to exhibit half-filled stripe order at very low temperatures   \cite{PhysRevX.11.031007}. 
Our calculations suggest that, in the pure Hubbard model, the half-filled stripe state appears to be special to width-4 cylinders, and we see filled stripes become the ground state in wider cylinders.
In Fig.~\ref{1_12_4_8}, we show an example of $\delta=1/12$ and $U=12$, in four different simulation cells --- two with-4 cylinders, $24\times 4$ and $48\times 4$, and two width-8 cylinders
 $24\times 8$ and $48\times 8$. Each calculation is performed following the same procedure that we have outlined. We see that the 
half-filled stripe is indeed the ground state for width-4 cylinders upon convergence of the self-consistent AFQMC. For width-8 cylinders, however, 
the ground state corresponds to filled stripes. 
We also compare the computed energies of half-filled and filled stripes at $U = 12$ and $\delta = 1/12$ in Table.~\ref{E_4_8}. We find that the energy for half-filled stripe
is lower than that of the filled stripe in width-4 systems but this trend 
is reversed in width-8 systems.
These results indicate that  the half-filled stripes in width-$4$ cylinder are affected by finite size effects, and the stripes become filled at the TDL.

Recent studies in larger cells, for example using variational Monte Carlo \cite{10.21468/SciPostPhys.7.2.021}, have suggested that 
the spin order might show wavelengths of $\alpha/\delta$, where $\alpha$ is neither $1$ (half-filled) nor $2$ (filled), for instance displaying a metallic 
state with $\alpha$ being a fraction.
We searched in a few such cases but did not find a stripe state with fractional $\alpha$. 
Below we show an example 
at $U = 8$ and $\delta= 1/12$. We computed the energies using trial wave functions with several different wavelengths, without invoking 
the self-consistency loop in the constraint. 
(In these cases the QMC results turned out to stay with the same wavelength, indicating that such a state is close in energy to the true ground state, as 
seen from the results below.)
PBC is applied along both directions  and the pinning field is removed, in order to allow direct comparison of the energies. 
As seen in Table~\ref{E_6_8},  
in the 
 $48\times 6$ lattice, the energy from the 2/3 filled stripe state ($\lambda=8$) is slightly lower than filled stripe state ($\lambda=12$), and both state are lower than half-filled stripe state ($\lambda=6$),   which is consistent with the result obtained in  Ref.~\cite{10.21468/SciPostPhys.7.2.021}. 
 (Note that 
  the best variational wave function gives an energy that is $\sim 0.013t$ per site higher.)
 In the $48\times 8$ lattice,  the energies of the filled and 2/3 filled stripe state  are almost degenerate. 
  We next performed a calculation with a trial wave function constructed from a linear combination of all three states,  and the result is 
  a filled stripe state. We thus conclude that, to within our resolution, the ground state is a 
  filled stripe state.
 






\begin{table}[b]
\caption{Comparison of the computed energies per site
in the half-filled and filled stripe states at $U = 12$ and $\delta = 1/12$, in cylinders with widths $4$ and $8$, and lengths $24$ and $48$. 
The system setup is the same as in Fig.~\ref{1_12_4_8}.
Half-filled stripe state has lower energy in width-4 systems but higher energy in width-8 systems.
A correction has been applied to the energies to account for finite Trotter step size \cite{Zhang97}.
}
\begin{tabular}{  c  c  c  }
 \hline\hline
  lattice & half-filled stripe & filled stripe \\
  \hline
 $24\times4$ & -0.5639(1) & -0.5630(1)\\  
 \hline
 $48\times4$ & -0.5583(2) & -0.5569(2) \\
 \hline
 $24\times8$ & -0.5596(2) & -0.5613(2)\\  
 \hline
 $48\times8$ & -0.5532(2) & -0.5541(2) \\
 \hline\hline
\end{tabular}
\label{E_4_8}
\end{table}

\begin{table}[b]
\caption{Comparison of the energy per site 
in the half-filled ($\lambda=6$),  2/3 filled ($\lambda=8$) and filled  ($\lambda=12$) stripe states at $U = 8$ and  $\delta=1/12$ in two different lattice 
sizes.
Fully periodic supercells are studied here, with no pinning fields.
Half-filled stripe state has the highest energy in both systems. The 2/3-filled  stripe state has the lowest energy in the width-6 supercell, while
its energy is indistinguishable from that of the filled stripe state in the width-8 supercell.
Both self-consistency and the use of a linear combination of trial wavefunctions with different stripe fillings lead to the filled stripe state as the ground state.
}
\begin{tabular}{  c  c  c  c}
 \hline\hline
  lattice & $\lambda=6$ & $\lambda=8$  & $\lambda=12$\\
  \hline
 $48\times6$ & -0.6820(2) & -0.6862(2)  & -0.6855(2) \\  
 \hline
 $48\times8$ & -0.6821(2) & -0.6854(2)  &  -0.6852(2)\\
 \hline\hline
\end{tabular}
\label{E_6_8}
\end{table}

\section{Phase diagram}
\label{phase}

%
%
%

\begin{figure}
\includegraphics[width=8cm]{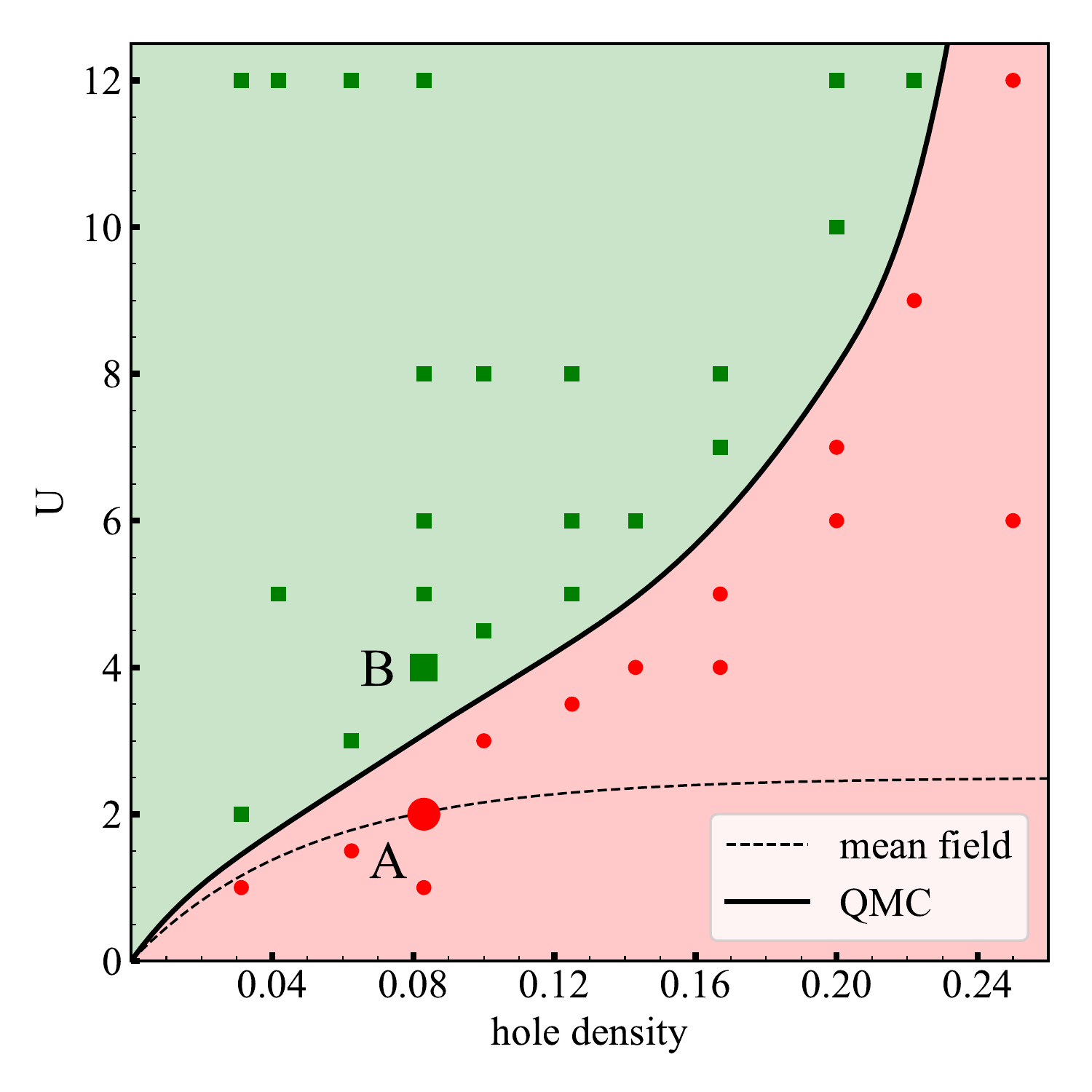}
\caption{Phase diagram of spin and charge orders in the pure Hubbard model. The black curve gives a rough estimation of the phase boundary based on the green squares representing parameters with modulated AFM spin and charge order and red cycles representing those without. The black dashed curve is the phase boundary from unrestricted Hartree-Fock \cite{Xu_2011} for reference.  
 A ($U=2$, $\delta=1/12$) and B ($U=4$, $\delta=1/12$) denote the two examples shown in Figs.~\ref{U_2_4_spin}, \ref{U_2_4_spin_struct}, and \ref{U_2_4_tdl}.
 }
\label{phase-diagram}
\end{figure}

Using the procedure described in the previous section, we map out the phase diagram of the spin and charge orders in the ground state 
of the pure two-dimensional Hubbard model ($t'=0$) in the TDL, from weak ($U\sim 0$) to fairly strong ($U\sim 12$). 
The results are summarized in  Fig.~\ref{phase-diagram}.
We find that, up to a doping value of $\delta\sim 0.2$, there exists a critical interaction strength $U_c(\delta)$, above which the system exhibits a collective 
mode of a modulated AFM order. The spin order has wavelength  $2/\delta$, and 
is accompanied by a charge order of wavelength $1/\delta$,
in which the hole density tends to be higher at the nodes of the spin order. 

At weaker $U$ (above $U_c(\delta)$) or larger doping, 
the charge order 
is weak. In these states the hole density is not vanishingly small away from the nodes of spin order; in fact the hole density remains substantial throughout space and is either constant or only shows a slow-varying wave with modest peaks at the nodes of the modulated AFM. We have referred to such states as SDWs (which can have charge order). 
As the interaction strength is increased and the doping is reduced, the SDW states evolve into stripe states, where the holes become more 
and more localized at the nodes. The distinction between the SDW and stripe states is not absolute, but it is important to emphasize that 
a modulated AFM order can exist with two different kinds of behaviors for the holes: mobile and wavelike vs.~localized and particle-like 
 \cite{PhysRevLett.104.116402}.

In Fig.~\ref{phase-diagram}, green squares represent parameters which lead to a ground state with SDW or stripe order in the TDL,  while red circles represent those which do not.   
Within Hartree-Fock diagonal stripes are found to be more stable than linear (along $x$- or $y$-direction) stripes
 at large $U$   \cite{Xu_2011}; 
diagonal stripes were also 
found to be close in energy with linear stripe state in the doped $t$-$J$ model \cite{PhysRevLett.113.046402,2020npjQM...5...28D}.
We searched for diagonal stripes in the Hubbard model in the parameter regime studied here, but did not find them to be the ground state.
(More details are given in the appendix.) 
Note that results from an inhomogeneous dynamical mean-field theory (iDMFT) study \cite{PhysRevB.89.155134} are in reasonable agreement with our results.

Based on our results, we show an estimate for the phase boundary as a solid black line in Fig.~\ref{phase-diagram},
whose position is not to be taken literally but which is bracketed 
by the data points around it. 
(In the appendix, we show two example scans, one fixed at $U=6$ varying $\delta$, and the other at $\delta=1/12$ varying $U$, to illustrate 
how the waves evolve across the transition line.) 

From the results, we see that the critical interaction strength $U_c(\delta)$ increase with
the doping level $\delta$. Nothing special is seen around the doping value of $\delta=1/8$.
The SDW or stripe order  persists from small doping near half-filling to doping levels as large as $1/5$. 

As discussed in more detail in Sec.~\ref{ssec:determine-lambda}, 
we did not find phase separation or non-filled SDW/stripe orders which survived in our finite-size scaling procedure.
This of course does not completely rule out such phases, because of the delicate nature of the different competing states and sensitivity to finite-size
and other effects, as well as possible systematic errors in the calculation. However, it does provide a rather stringent screening of other possible states, given the high accuracy and extensive nature of these calculations.

\section{conclusion and perspective}
\label{con}


Employing state-of-the-art AFQMC methods with self-consistent constraint 
and performing finite-size scaling to large simulation cell sizes, we map out the ground state phase diagram of the doped 2D Hubbard model regarding the spin and charge orders as a function of doping $\delta$ and interaction strength $U$.
Modulated SDW or stripe orders are found to exist for doping as large as $\sim 1/5$ with sufficiently large $U$.
The period of the spin (charge) density wave was found to be $2/\delta$ ($1/\delta$), which implies that the ground state stripes (at larger $U$) are filled.
Our results show that stripe/SDW
 exists not only in the vicinity of $1/8$ doping, but also extends to very small doping near half-filling and to the overdoped region. 
 Recent experiments in cuprate \cite{Miao_21} found that stripe order exists with doping beyond the superconducting dome, remaining observable for as large as $\delta \approx 0.21$. In the future, it will be interesting to study how the phase
 diagram changes with the inclusion of a small next-nearest hoping $t^\prime$. The inclusion of $t^\prime$ frustrates the antiferromagnetic order at half-filling and we anticipate that it will cause
the critical interaction $U_c$ to increase for a given doping level, and potentially change the properties of the orders. Superconductivity is found to be absent in the pure Hubbard model \cite{PhysRevX.10.031016} and very recent
results indicate 
that a small positive $t^\prime$ can induce superconductivity in the doped $t$-$J$ model  \cite{2021arXiv210403758G,2021arXiv210410149J,PhysRevLett.127.097002}. The relationship between stripe and superconductivity
is an important topic for future investigation.

\section{ACKNOWLEDGEMENTS}

We thank  Yuanyao He, A.~Georges, A.~J.~Millis, S.~R.~White for helpful discussions. 
H.X.~gratefully acknowledge the Center for Computational Quantum Physics (CCQ), Flatiron Institute for support and hospitality.
M.Q.~is supported by a start-up fund from School of Physics and Astronomy in Shanghai Jiao Tong University. 
Most of the computing was carried out at the Flatiron Institute, with the rest carried out at computational facilities of College of William and Mary,  and XSEDE. The Flatiron institute is a division of the Simons Foundation.

\bibliography{sdw_cite}

%
%
%
\appendix 

\section{Filling of stripes}
In Fig.10,  we show additional results for the spin and hole densities in three other systems to supplement Fig.~\ref{1_12}:
  $\delta=1/10$ and $U=5$,  $\delta=1/8$ and $U=6$, and $\delta=1/6$ and $U=8$.

\begin{figure*}
   \centering
       \includegraphics[width=55mm]{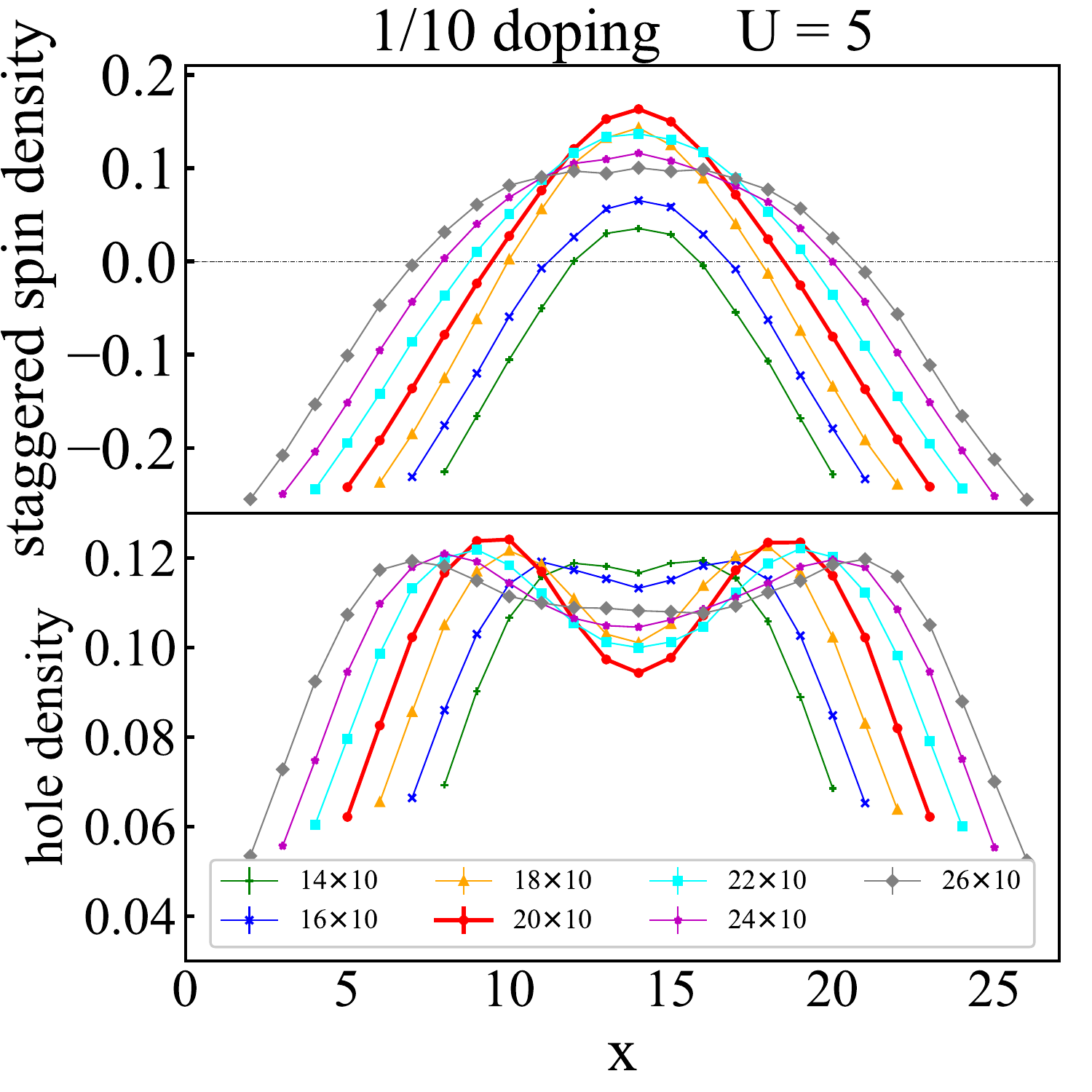}
       \includegraphics[width=55mm]{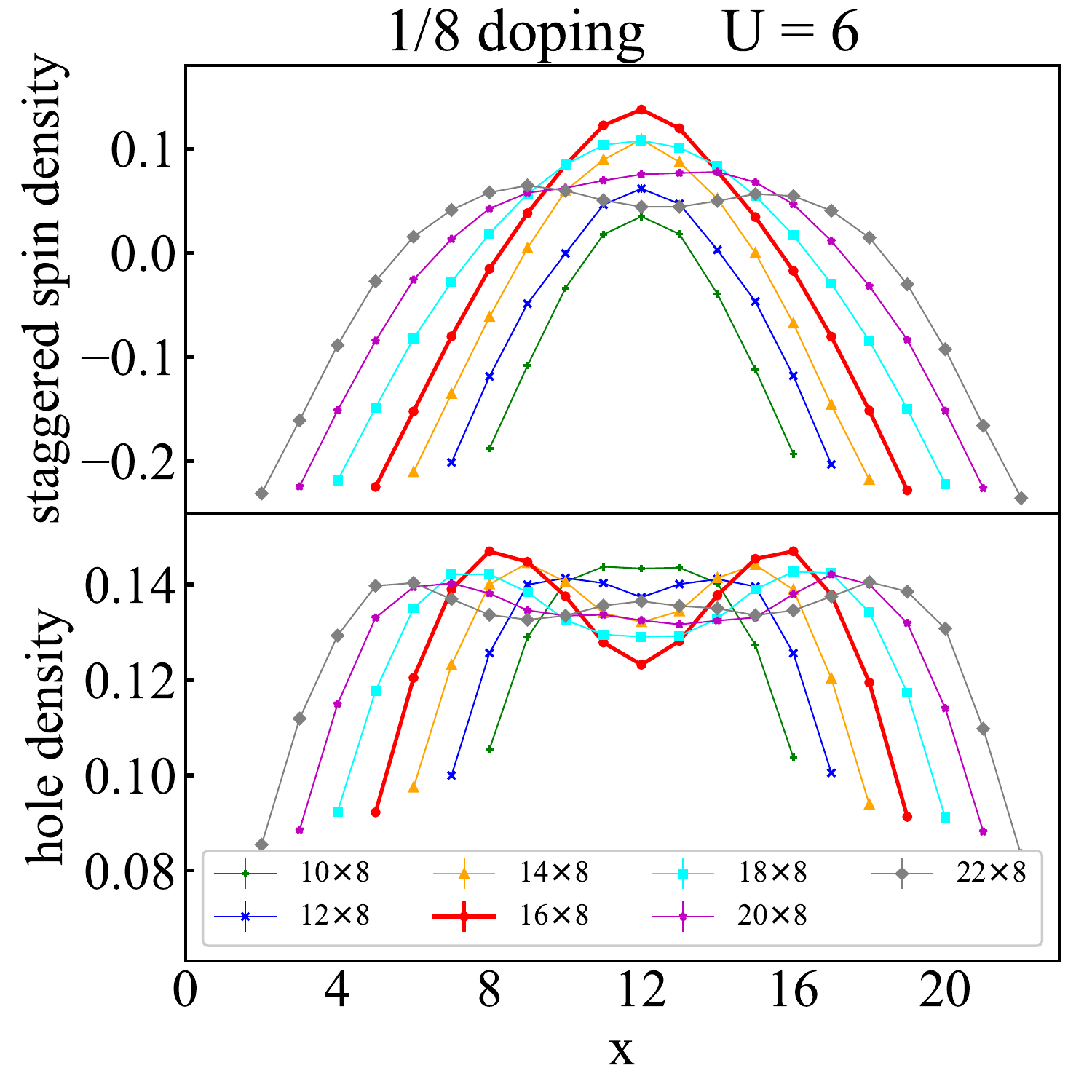}
       \includegraphics[width=55mm]{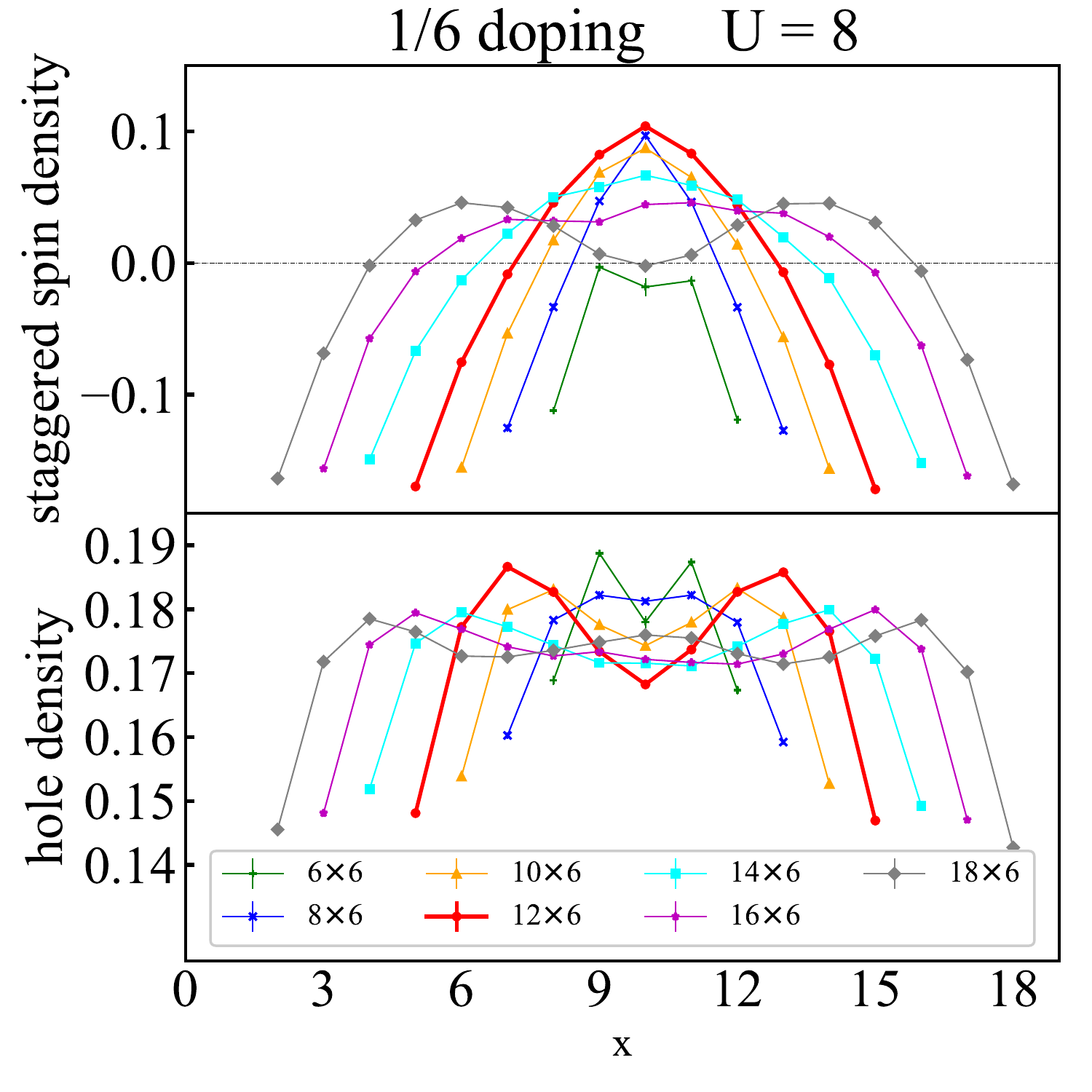}

      \caption{Similar as Fig.~\ref{1_12}. The staggered spin and charge densities are shown for $1/10$ doping ($U = 5$), $1/8$ doping ($U = 6$) and $1/6$ doping (U = $8$).}
      \label{app_1}

\end{figure*}

\section{Examples of parameter scans in the phase diagram}

\begin{figure}
    \includegraphics[width=84mm]{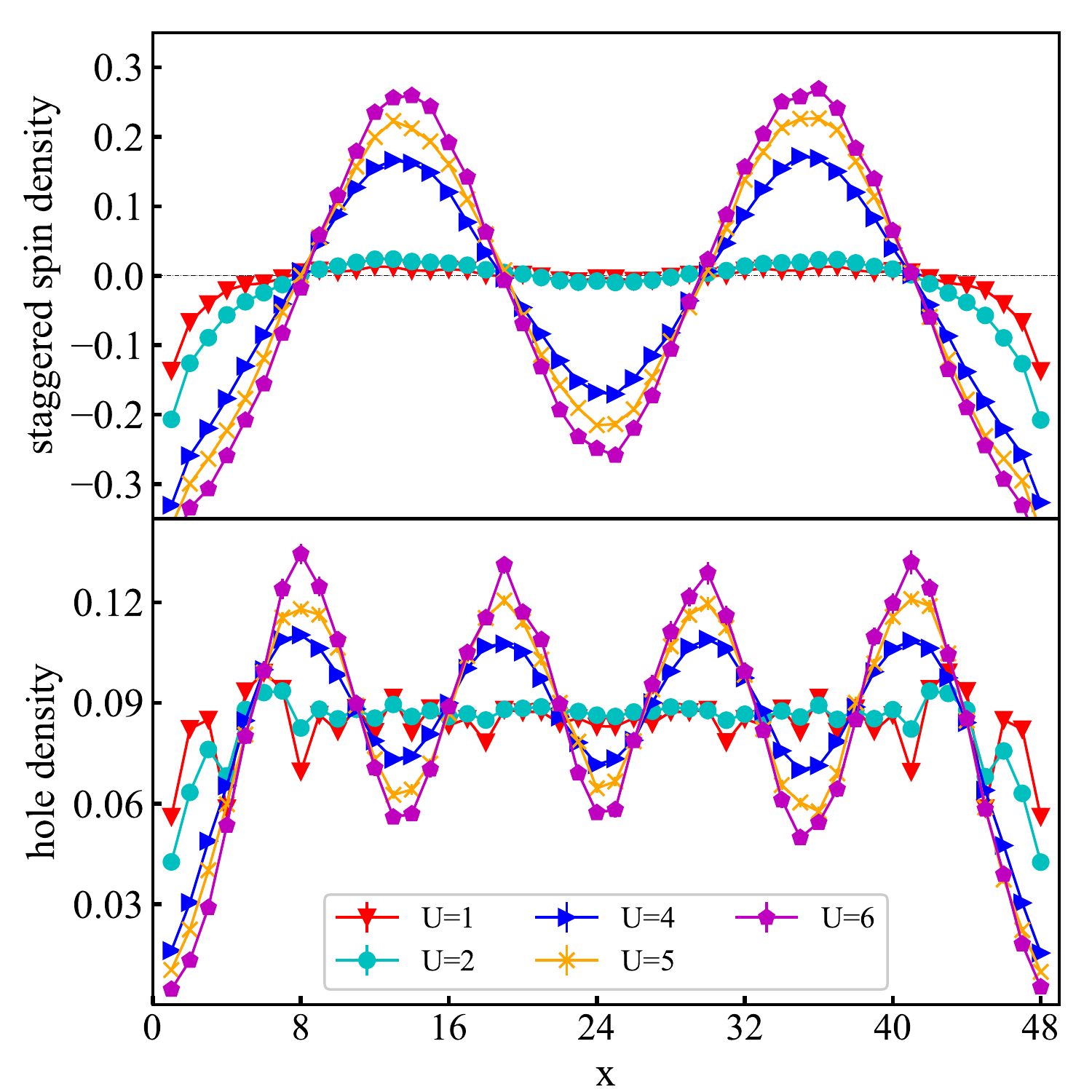}

\caption{Staggered spin (up) and hole (down) density at $1/12$ doping for different $U$ values. The stripe order develops with $U \ge 4$.}
\label{U_dependence}
\end{figure}

\begin{figure}

\includegraphics[width=84mm]{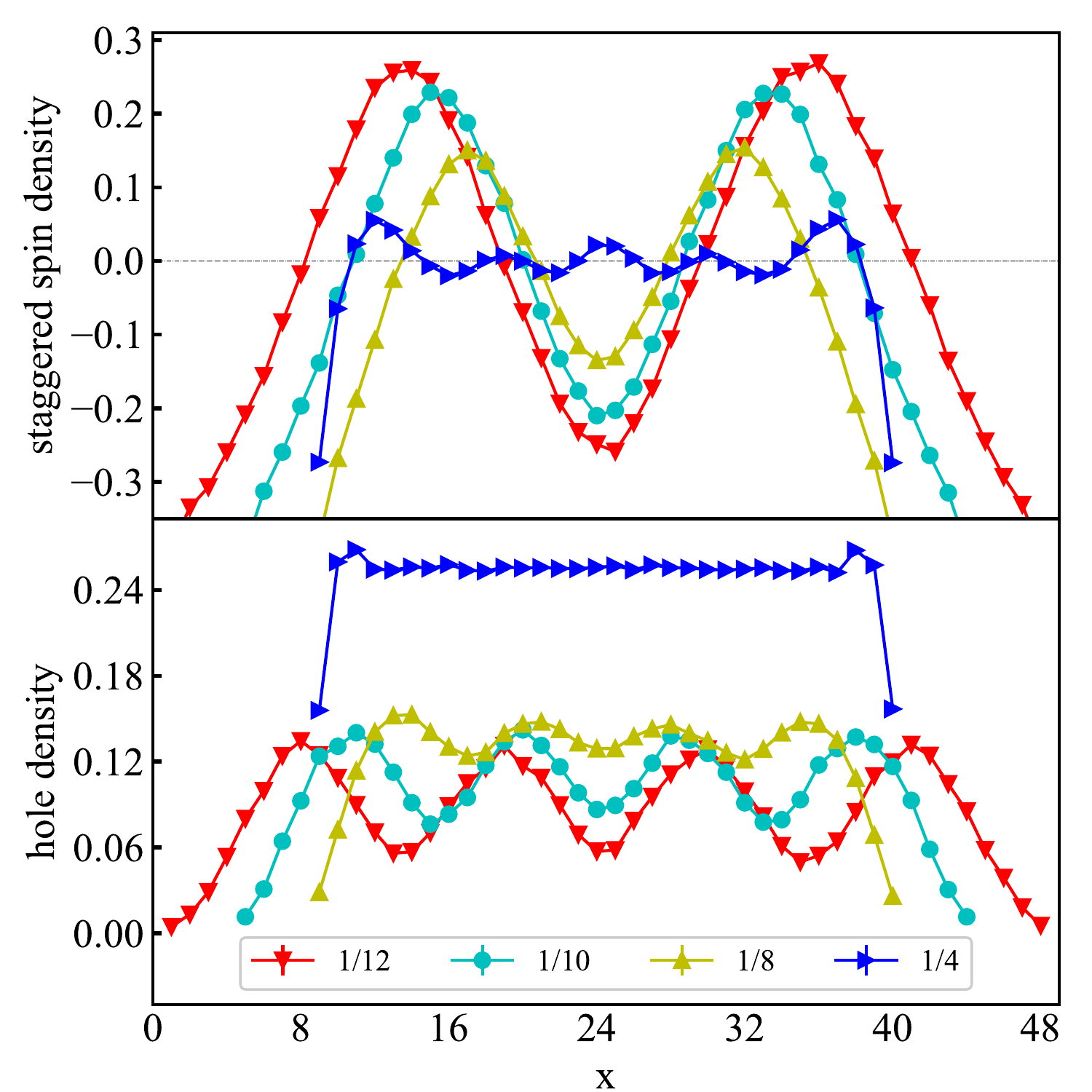}

\caption{Staggered spin (up) and hole (down) density at $U = 6$ for different dopings. Order only develops with $\delta \le 1/8$.}
\label{doping_dependence}
\end{figure}

In Fig.~\ref{U_dependence}, we show how the spin and charge orders evolve as a function of $U$ at fixed doping $\delta = 1/12$. The system is a cylinder with size $8 \times 48$. 
A modulated AFM order develops only when $U \ge 4$.
%
%
Similarly, in Fig.~\ref{doping_dependence}, we show the results of a scan  
at fixed interaction strength $U = 6$, in systems of size $L_x\times 6$, with $L_x$ 
 from 32 to 48 to accommodate  $1/4$, $1/8$ $1/10$, and $1/12$ doping. 
As can be seen, no order is present until $\delta \le 1/8$.

\section{Diagonal vs.~linear stripes}

In Hartree-Fock calculations, stripe states in the diagonal direction were observed at small doping and large $U$ \cite{Xu_2011}.
Studies using dynamical mean-field theory (DMFT) with exact diagonalization solvers also predicted  
 diagonal stripes for $\delta < 0.05$ \cite{fleck1};
In Table.~\ref{E_diag},
we present energy comparison for linear and diagonal stripe state at $U = 12$ for $1/16$, $1/24$, and $1/32$ dopings.
The energies are calculated with TABC to minimize finite size effects \cite{PhysRevE.64.016702}. As we can see in Table.~\ref{E_diag}, in our calculations the linear stripe order always has a lower energy than the diagonal stripe for doping as low as $\delta = 1/32$.
Our systematic calculations at  $\delta = 1/32$, $U=2$ and $\delta = 1/24$, $U=5$ also yielded a linear SDW or stripe state.


\begin{table}[t]
\begin{tabular}{ | c | c | c | c |}
 \hline
  doping & 1/16 & 1/24 &  1/32   \\
  \hline 
  lattice  &  $32\times 8$  &  $48\times6$ & $64\times6$   \\  
 \hline
 linear stripe & -0.5095(2) & -0.4611(1) & -0.4371(2) \\
 \hline
  diagonal stripe & -0.5076(1) & -0.4595(1) & -0.4362(1)\\
  \hline
\end{tabular}
\caption{Energy (per site) comparison for linear and diagonal stripe states at $U=12$. The linear stripe state consistently has lower energy.}
\label{E_diag}
\end{table}

\end{document}